\documentclass[aps,pra,reprint,twocolumn,showpacs,showkeys,floatfix,nobibnotes]{revtex4-2}

\usepackage{hyperref}
\hypersetup{colorlinks=true, urlcolor=red, citecolor=blue, linkcolor=blue}
\usepackage{graphicx}
\usepackage{color}
\usepackage{subfigure}
\usepackage{epstopdf}
\usepackage{lastpage,epstopdf}

\begin{document}
\title{Nonclassicality of photon-added-then-subtracted and photon-subtracted-then-added states}
\author{Arpita Chatterjee}
\email{mailtoarpita@rediffmail.com}
\affiliation{School of Physical Sciences, Jawaharlal Nehru
University, New Delhi 110067, India}
\date{\today}

\begin{abstract}
\begin{center}
\end{center}
We formulate the density matrices of a quantum state obtained by first adding multi-photons to and then subtracting multi-photons from any arbitrary state as well as performing the same process in the reverse order. Considering the field to be initially in a thermal (or in an even coherent) state, we evaluate the photon number distribution, Wigner function and Mandel's $Q$ parameter of the resulting field. We show graphically that in which order multi-photons are added and subtracted has a noticeable effect on the temporal behavior of these statistical properties.
\end{abstract}

\pacs{42.50.-p, 42.50.Ct, 42.50.Pq}

\maketitle


\section{Introduction}

The non-commutativity between the annihilation ($a$) and creation ($a^\dag$) operators has long been a field of interest in quantum
mechanics. Due to this non-commutativity of bosonic operators, simple alternated sequences of adding and subtracting identical
particles to any quantum system show different results. Agarwal and Tara \cite{agarwal91} first proposed a method for producing
the photon-added coherent state. Another way of creating photon-added or photon-subtracted state is through a beam-splitter
\cite{ban96}. Dakna \cite{dakna98} showed that if any arbitrary initial state and a Fock state are injected at the two input channels,
then the photon number counting of the output Fock state reduces the other output channel to a corresponding photon-added or
photon-subtracted state. In addition, a cavity-QED based technique was theoretically discussed by Sun \textit{et al.} \cite{sun08}.
Conditioned on sending two atoms one by one at the considered levels and detecting them only if they end at the desired levels, they
verified that $a$ and $a^\dag$ are non-commutable.

Recently, Parigi \textit{et al.} \cite{parigi07} successfully demonstrated an experimental set-up to observe the effect of adding or subtracting
single-photon to or from a completely classical and fully incoherent thermal light field. By applying alternated sequences of the creation and annihilation operators they realized that the resulting states depend on the order in which the two quantum operators have been applied. The same group also implemented a single-photon interferometer to achieve a direct proof for the bosonic commutation relation \cite{kim08}. Besides testing the non-commutativity of bosonic operators, this adding-subtracting phenomena has a great significance because excitation by a definite number of photons can turn any classical field to a nonclassical one \cite{jones97}. On the other hand, annihilation of a quantum state not only produces nonclassicality but it is able to convert Gaussian states into non-Gaussian ones \cite{barbieri10}. Non-Gaussian states are known to provide useful resources for tasks such as entanglement distillation \cite{ourjoumtsev07} and noiseless amplification \cite{xiang10}. In recent times, successful experiments have been proposed by Zavatta \textit{et al.} \cite{zavatta04,zavatta05,zavatta07} to manipulate photon subtraction from or photon addition to a light beam via simple optical processes such as beam splitter, frequency down-conversion and homodyne detection. These experimental successes has made possible the generation of nonclassical states which have many real life applications. For example, squeezed states are used to reduce the noise level in one of the phase-space quadratures below the quantum limit \cite{scully97}, entangled states are employed to realize quantum computer and to transfer quantum information \cite{braunstein05} etc. In fact, besides performing quantum communication, it has been experimentally shown that entanglement can be enhanced by subtracting a photon from one of the two modes of a two-mode squeezed state \cite{ourjoumtsev07}. Here we concentrate on the behavioral changes of nonclassicality of quantum states after applying multi-photons in different orders.

It is interesting to notice that the theoreticians as well as the experimentalists investigated a lot about single or two photon addition
or subtraction but less attention has been paid to multi-photon changes. For example, Marek \textit{et al.} \cite{marek08} generated the squeezed superpositions of coherent states by applying the two-photon subtraction ($a^2$) or the photon subtraction and addition ($a^\dag a$) combination to squeezed vacuum state. But in our paper, we are interested in finding the results of using an arbitrary $p$-photon addition and $q$-photon subtraction. Our multi-photon scheme can be realizable in a quantum optics laboratory as the initial thermal (even coherent) field contains a very small number of photons \cite{kim41}. Yang Yang \textit{et al.} \cite{yang09} investigated the nonclassicality of a single-photon-subtracted Gaussian state as well as a photon-added-then-subtracted thermal state. They used nonclassical depth as a measure of nonclassicality and observed a strong correlation between the nonclassicality of the radiation field and the photon addition-subtraction process.  They reported that the states generated by first adding (subtracting) multi-photons to an arbitrary state and then subtracting (adding) multi-photons from the resulting state is certainly nonclassical if the number of added photons is equal to or larger than the number of subtracted photons. It has been pointed out that the photon-added-then-subtracted state is nonclassical irrespective of any initial state. But their approach is restricted to the nonclassicality depth criteria \cite{lee91} only. We here employ other popular tools for looking into the nonclassicality of multi-photon cases.

This paper is structured as follows: we describe the density matrices for added-then-subtracted and subtracted-then-added quantum states in Sec.~\ref{sec2}. Sec.~\ref{sec3} concerns with finding various distributions of the thermal field after photon excitation and de-excitation processes. In Sec.~\ref{sec4}, we study the same properties for the even coherent state. The last section ends with a summary of the main results of this article.

\section{General Theory}
\label{sec2}

Here we do a comparison for the nonclassicality of added-then-subtracted and subtracted-then-added state. The density
matrix of an arbitrary quantum state of the single-mode radiation field can be expanded in terms of photon number states as
\begin{eqnarray}
\hat{\rho} = \sum_{m=0}^\infty \sum_{n=0}^\infty \rho(m,
n)|m\rangle\langle n|.
\label{eq1}
\end{eqnarray}
For a given density matrix $\hat{\rho}$ of the single-mode radiation field, the state generated by first adding $p$ photons
and then subtracting $q$ photons may be written as
\begin{eqnarray}
\hat{\rho}^{(sa)}=N_1 a^q a^{\dag p} \hat{\rho} a^p a^{\dag q},
\label{eq2}
\end{eqnarray}
where $N_1$ is the normalization constant for the density operator. Substituting (\ref{eq1}) into (\ref{eq2}), we obtain
\begin{eqnarray}\nonumber
\hat{\rho}^{(sa)} & = & N_1\sum_{m=0}^\infty \sum_{n=0}^\infty
\rho(m, n)\frac{(m+p)!}{\sqrt{m!}\sqrt{(m+p-q)!}}\\
& & \times|m+p-q\rangle\langle n+p-q|\frac{(n+p)!}{\sqrt{n!}\sqrt{(n+p-q)!}}.
\label{eq3}
\end{eqnarray}
Next we consider just the reverse process of Eq.~(\ref{eq2}), i.e., first subtracting $q$ photons and then adding $p$ photons to the initial
state. The finally generated state is
\begin{eqnarray}\nonumber
\hat{\rho}^{(as)} & = & N_2\sum_{m=q}^\infty \sum_{n=q}^\infty
\rho(m,
n)\frac{\sqrt{m!}\sqrt{(m+p-q)!}}{(m-q)!}\\
& & \times|m+p-q\rangle\langle n+p-q|\frac{\sqrt{n!}\sqrt{(n+p-q)!}}{(n-q)!}.
\label{eq4}
\end{eqnarray}
We can derive any property of the final field from these density operators.

\section{Thermal state}
\label{sec3}

For the initial thermal field the density operator is \cite{sun08}
\begin{eqnarray}
\hat{\rho}_{\text{th}} = \sum_{n=0}^\infty
\frac{{\bar{n}}^n}{(1+\bar{n})^{1+n}}|n\rangle\langle
n|,\label{eq5}
\end{eqnarray}
where $\bar{n}$ is the mean photon number of the thermal state. Using the formulas (\ref{eq2}) and (\ref{eq3}) the final density operators
for the two sequences are
\begin{eqnarray}\nonumber
\hat{\rho}_{\text{th}}^{(sa)} & = & N_1\sum_{n=0}^\infty
\frac{{\bar{n}}^n}{(1+\bar{n})^{1+n}}
\frac{((n+p)!)^2}{n!(n+p-q)!}\\
& & \times|n+p-q\rangle\langle n+p-q|,\label{eq6}
\end{eqnarray}
and
\begin{eqnarray}\nonumber
\hat{\rho}_{\text{th}}^{(as)} & = & N_2\sum_{n=q}^\infty
\frac{{\bar{n}}^n}{(1+\bar{n})^{1+n}}
\frac{n!(n+p-q)!}{((n-q)!)^2}\\
& & \times|n+p-q\rangle\langle n+p-q|.\label{eq7}
\end{eqnarray}
The analytical expressions for $N_1$ and $N_2$ are respectively

\begin{equation}
N_1 =
\left\{
\begin{array}{lcl}
\frac{(1+\bar{n})(p-q)!}{(p!)^2~{_2}F_1(1+p,1+p;1+p-q;\frac{\bar{n}}{1+\bar{n}})},~~\text{$p-q \geq 0$}\\\\
\frac{(1+\bar{n})}{\sum_{n=0}^\infty (\frac{\bar{n}}{1+\bar{n}})^n\frac{((n+p)!)^2}{n!(n+p-q)!}}~~~~~~~~,~~\text{$p-q<0$}
\end{array}
\right.
\label{eq8}
\end{equation}
and
\begin{eqnarray}
N_2 = \frac{(1+\bar{n}){\left(\frac{\bar{n}}{1+\bar{n}}\right)}^{-q}}{p!~q!~{_2}F_1(1+q,1+p;1;\frac{\bar{n}}{1+\bar{n}})},
\label{eq9}
\end{eqnarray}
in which ${_P}F_Q$ is the Generalized Hypergeometric function.

\subsection{Photon Number Distribution}

For a given thermal field, the probabilities of finding $n$ photons in states (\ref{eq6}) and (\ref{eq7}) are respectively
\begin{eqnarray}
p_{\text{th}}^{(sa)}(n) = \frac{N_1}{(1+\bar{n})}\left(\frac{\bar{n}}{1+\bar{n}}\right)^{n-p+q}\frac{((n+q)!)^2}{n!(n-p+q)!},
\label{eq10}
\end{eqnarray}
and
\begin{eqnarray}
p_{\text{th}}^{(as)}(n) = \frac{N_2}{(1+\bar{n})}\left(\frac{\bar{n}}{1+\bar{n}}\right)^{n-p+q}\frac{n!(n-p+q)!}{((n-p)!)^2}.
\label{eq11}
\end{eqnarray}

\begin{figure*}[ht]
\centering
\includegraphics[width=5cm]{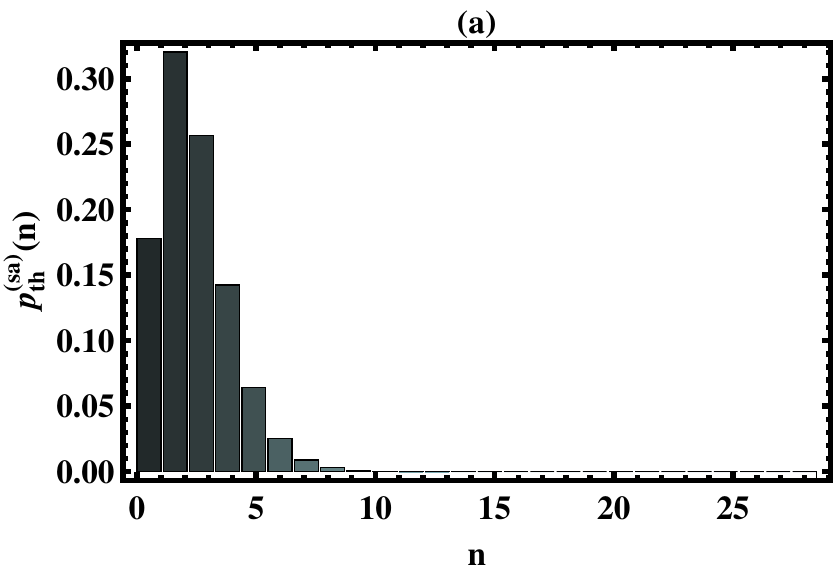}\hspace{1cm}
\includegraphics[width=5cm]{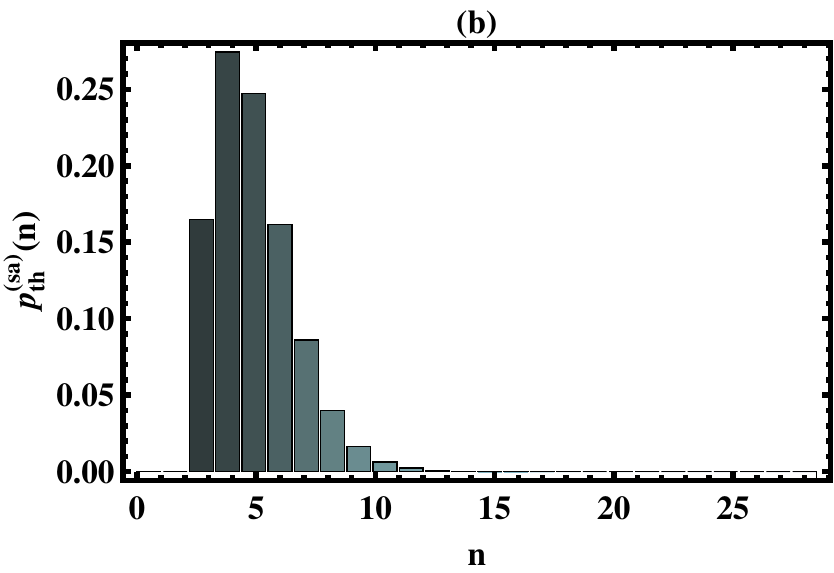}\hspace{1cm}
\includegraphics[width=5cm]{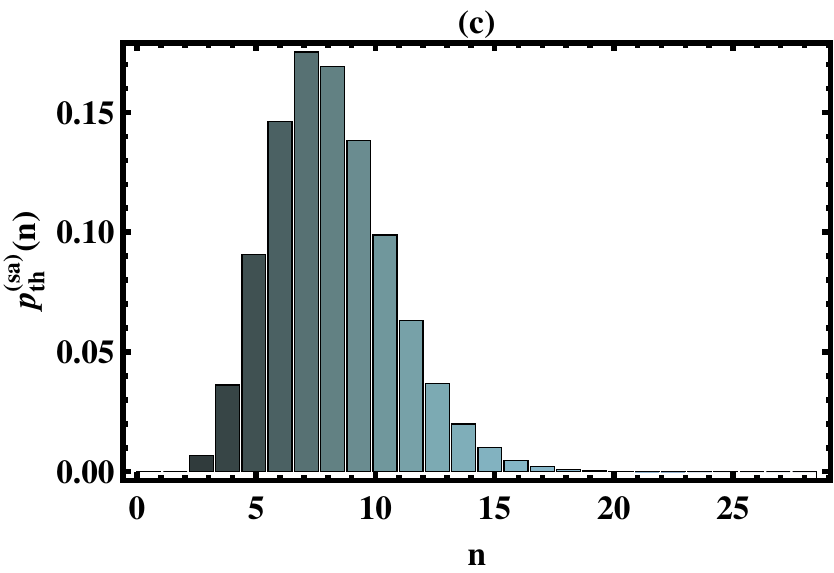}\hspace{1cm}
\includegraphics[width=5cm]{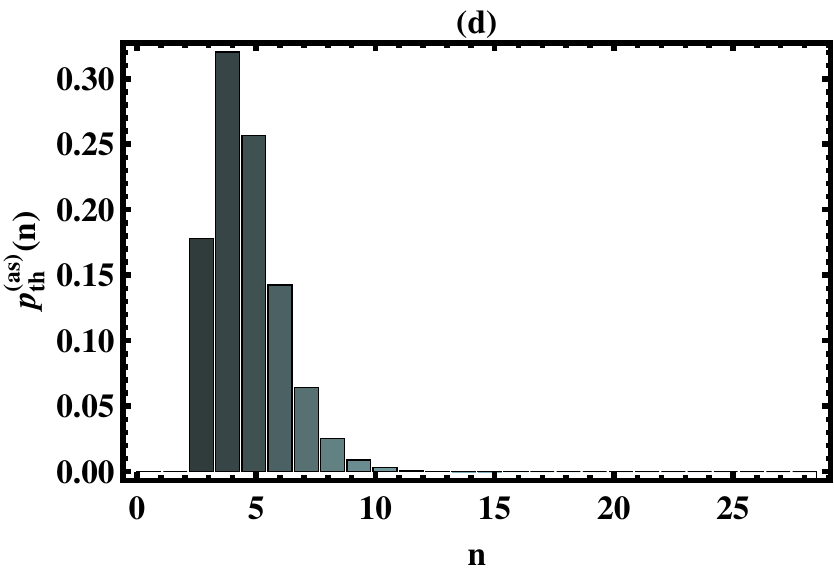}\hspace{1cm}
\includegraphics[width=5cm]{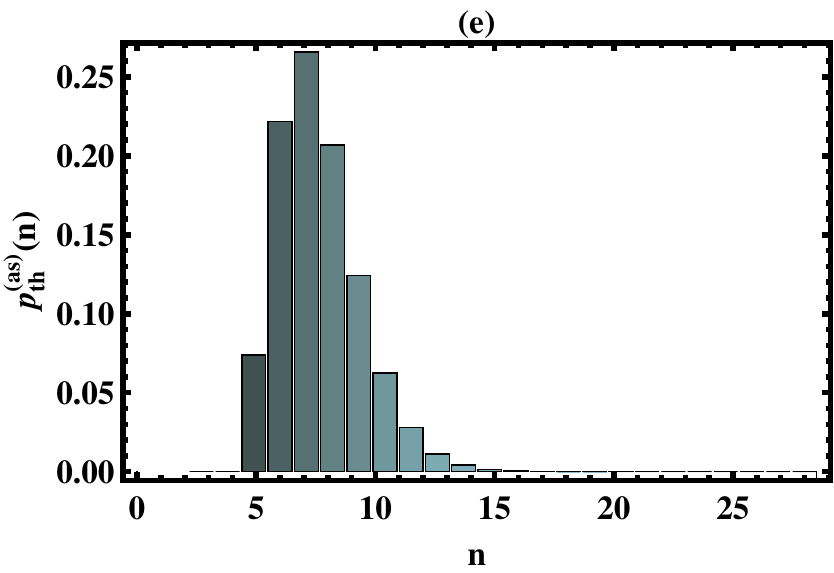}\hspace{1cm}
\includegraphics[width=5cm]{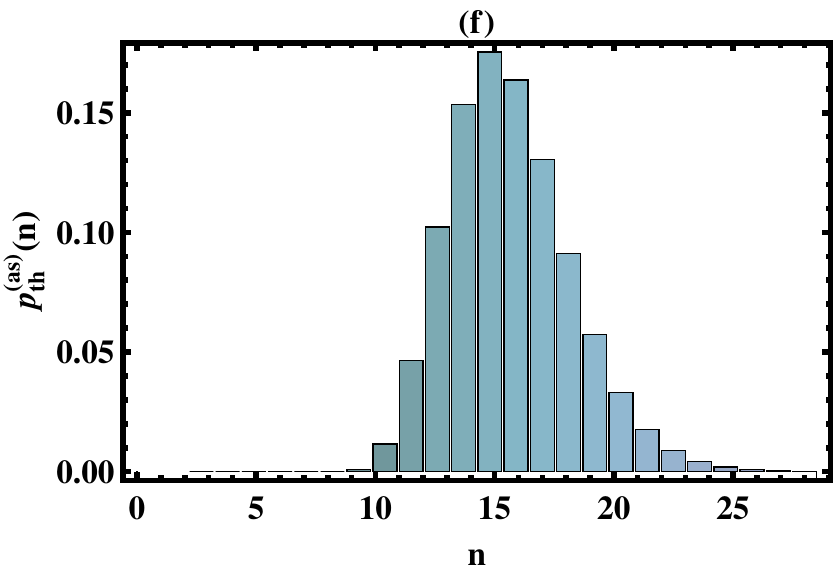}

\caption{(Color online) Photon number distribution of photon-added-then-subtracted (upper line) and photon-subtracted-then-added (lower line) thermal field is plotted against $n$ for $\bar{n}=0.25$ and (a) $p=q=2$, (b) $p=4$, $q=2$, (c) $p=8$, $q=6$, (d) $p=q=2$, (e) $p=4$, $q=2$ and (f) $p=8$, $q=6$.}
\label{fig1}
\end{figure*}

In Fig.~\ref{fig1}, we show how photon number distribution changes with photon number $n$ for different excitation and de-excitation parameters. In general, with increasing $p$ and $q$ the peak moves towards right and becomes more wide for both the added-then-subtracted and subtracted-then-added distributions. That means adding and subtracting photons shift the peak from zero to nonzero photons. We further notice that $p_{\text{th}}^{(as)}(n)$ possesses a narrower distribution compared to $p_{\text{th}}^{(sa)}(n)$ [see Figs.~\ref{fig1}(a)--(c) and Figs.~\ref{fig1}(d)--(f) ].

\subsection{Wigner Distribution}

For an optical field in the state $\hat{\rho}$, the Wigner function is defined as \cite{pathak05}
\begin{eqnarray}\nonumber
& & W(\beta, \beta^*)\\ & = & \frac{2}{\pi^2}e^{2|\beta|^2}
\int {\langle-\gamma|\hat{\rho}|\gamma\rangle
\exp[-2(\beta^*\gamma-\beta\gamma^*)]d^2\gamma},
\label{eq12}
\end{eqnarray}
where $|\gamma\rangle$ is a coherent state. In particular, a simple calculation via (\ref{eq12}) results the Wigner distribution for the initial thermal state as
\begin{eqnarray}
W_{\text{th}}(\beta, \beta^*) =
\frac{2}{\pi(1+2\bar{n})}\exp{\left(-\frac{2|\beta|^2}{1+2\bar{n}}\right)},
\label{eq13}
\end{eqnarray}
which is clearly Gaussian. The Wigner functions for photon-added-then-subtracted and photon-subtracted-then-added thermal states
are respectively
\begin{eqnarray}\nonumber
W_{\text{th}}^{(sa)}(\beta, \beta^*) & = &
\frac{2N_1}{\pi}e^{-2|\beta|^2}
\frac{(4|\beta|^2)^{p-q}}{(1+\bar{n})}\sum_{n=0}^\infty
\left\{\frac{(n+p)!}{(n+p-q)!}\right\}^2\\& &
\times\frac{\left\{\left(\frac{\bar{n}}{1+\bar{n}}\right)(4|\beta|
^2)\right\}^n}{n!},\label{eq14}
\end{eqnarray}
and
\begin{eqnarray}\nonumber
W_{\text{th}}^{(as)}(\beta, \beta^*) & = &
\frac{2N_2}{\pi}e^{-2|\beta|^2}
\frac{(4|\beta|^2)^{p-q}}{(1+\bar{n})} \sum_{n=q}^\infty
\left\{\frac{n!}
{(n-q)!}\right\}^2\\
& &
\times\frac{\left\{\left(\frac{\bar{n}}{1+\bar{n}}\right)(4|\beta|
^2)\right\}^n}{n!}.\label{eq15}
\end{eqnarray}

\begin{figure*}[ht]
\centering
\includegraphics[width=5cm]{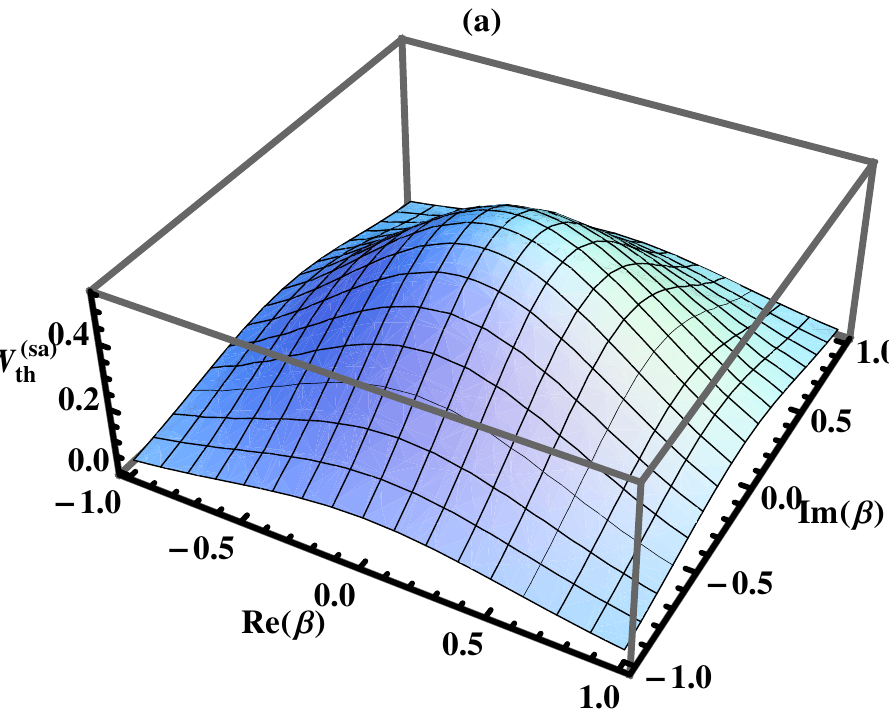}\hspace{1cm}
\includegraphics[width=5cm]{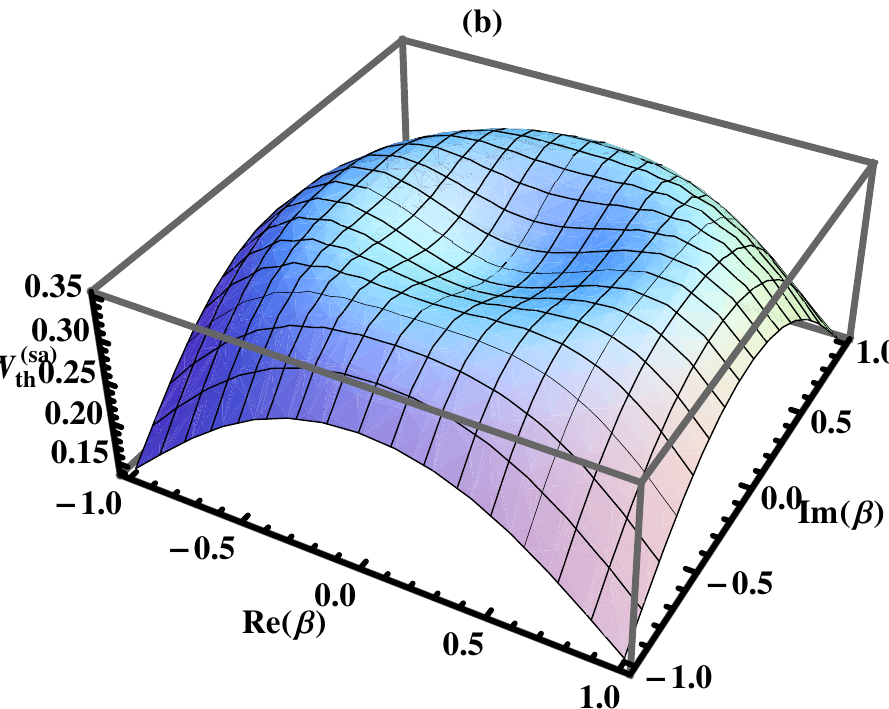}\hspace{1cm}
\includegraphics[width=5cm]{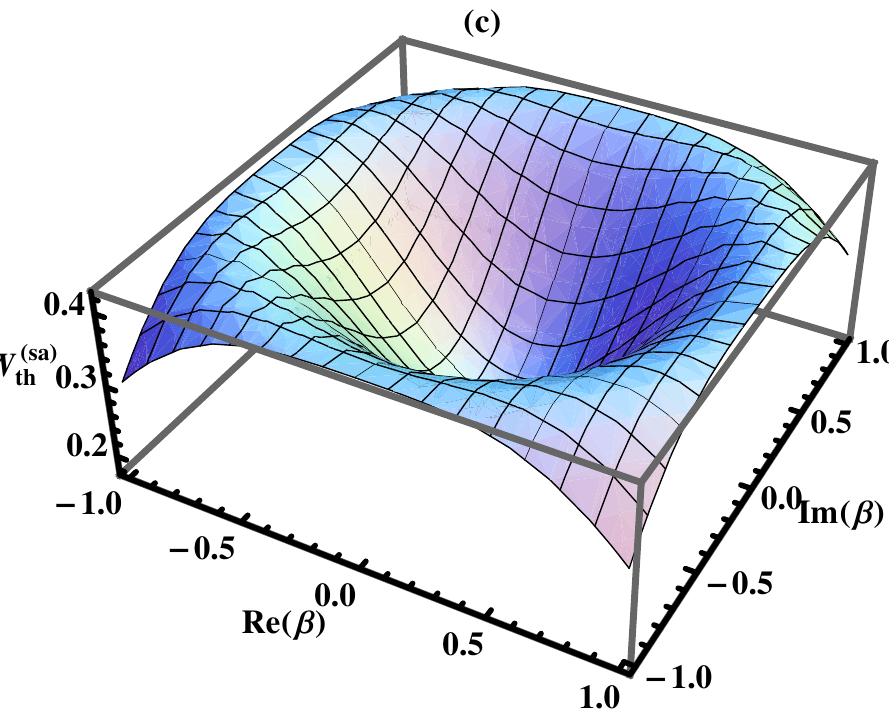}\hspace{1cm}
\includegraphics[width=5cm]{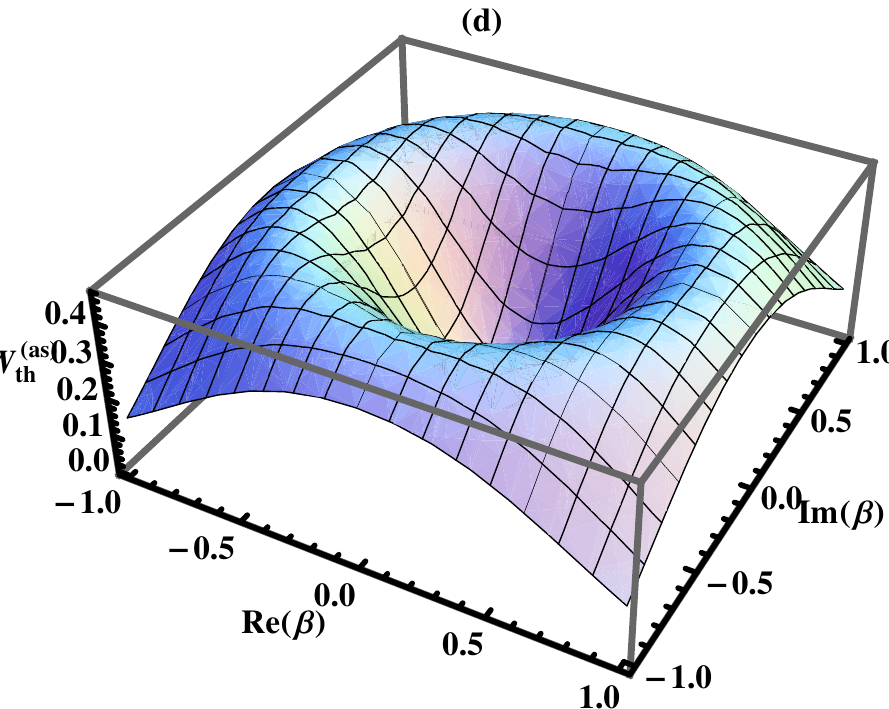}\hspace{1cm}
\includegraphics[width=5cm]{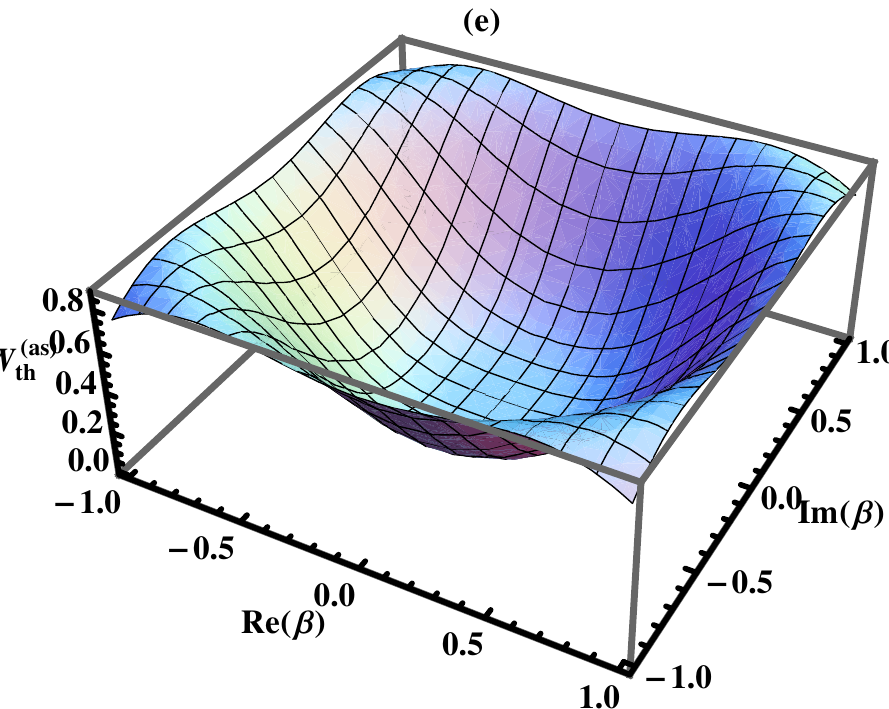}\hspace{1cm}
\includegraphics[width=5cm]{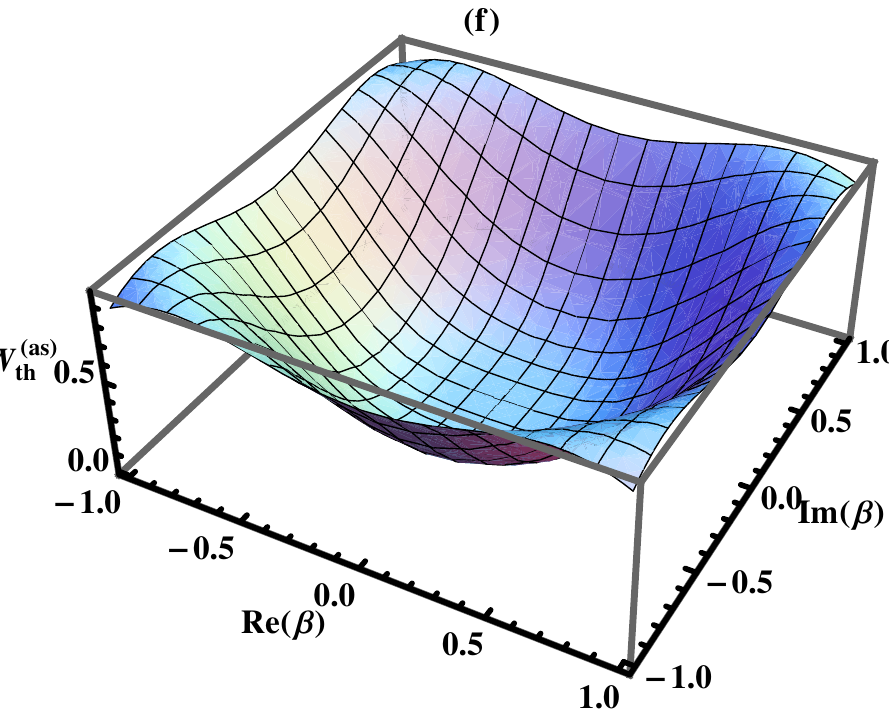}

\caption{(Color online) Wigner distribution of photon-added-then-subtracted (upper line) and photon-subtracted-then-added (lower line) thermal state as a function of Re$(\beta)$ and Im$(\beta)$ for $\bar{n}=0.04$ and (a) $p=q=1$, (b) $p=4$, $q=12$, (c) $p=8$, $q=12$, (d) $p=q=1$, (e) $p=2$, $q=4$ and (f) $p=2$, $q=6$.}
\label{fig2}
\end{figure*}

Fig.~\ref{fig2} elaborates the Wigner distributions in phase space for several combinations of $p$ and $q$. It is clear that the cycling of photons in alternate orders cause extensively different changes to the field. $W_{\text{th}}^{(sa)}(\beta, \beta^*)$ is positive everywhere but the Gaussian peak gradually transforms to a central dip as $(p, q)$ increases. Here addition of a larger number of photons, keeping $q$ fixed, leads to a deeper region of the Wigner function [see Figs.~\ref{fig2}(b)-(c)]. That means adding more photons in photon-added-then-subtracted method can prepare a classical non-Gaussian state \cite{allevi10}. In Figs.~\ref{fig2}(d)-(f), $W_{\text{th}}^{(as)}(\beta, \beta^*)$ are plotted for (1,1), (2,4) and (2,6) respectively. For this reverse process, the dip at the central position shrinks with increasing $p$ and $q$. When $p$ is fixed, the midway hole looses its depth as annihilation number increases. It should be noted that the Wigner function obtained by first adding one photon and then subtracting one photon [Fig.~\ref{fig2}(a)] from an initial thermal state remarkably differs in character from the Wigner function after the one-photon-subtracted-then-added process [Fig.~\ref{fig2}(d)]. The different results for these two alternate sequences with same ($p$,~$q$) establish the non-commutativity between $a$ and $a^\dag$.

\subsection{Mandel's $Q$ Parameter}

Next to determine the photon statistics of a single-mode radiation field we consider the Mandel's $Q$
parameter defined by \cite{mandel79}
\begin{eqnarray}
Q = \frac{\langle{a^\dag}^2 a^2\rangle}{\langle a^\dag a\rangle}-\langle a^\dag a\rangle,
\label{eq16}
\end{eqnarray}
which measures the deviation of the variance of the photon number distribution of the considered state from the Poissonian distribution of the coherent state.
$Q=0$ stands for Poissonian distribution, while for $-1\leq Q<0~(Q>0)$, the field obeys sub-~(super-)~Poissonian photon statistics. But the negativity of $Q$ is not a necessary condition to distinguish the quantum states into nonclassical and classical regime but just a sufficient one. For example, a state may be nonclassical even though $Q$ is positive \cite{agarwal91}.
Using (\ref{eq16}), one can easily calculate
\begin{eqnarray}\nonumber
Q_{\text{th}}^{(sa)} & = &
\frac{(p-q-1)~{_2}F_1(1+p,1+p;p-q-1;\frac{\bar{n}}{1+\bar{n}})}
{{_2}F_1(1+p,1+p;p-q;\frac{\bar{n}}{1+\bar{n}})}\\
& & -\frac{(p-q)~{_2}F_1(1+p,1+p;p-q;\frac{\bar{n}}{1+\bar{n}})}
{{_2}F_1(1+p,1+p;1+p-q;\frac{\bar{n}}{1+\bar{n}})}
\end{eqnarray}
and
\begin{eqnarray}\nonumber
Q_{\text{th}}^{(as)} & = &
\frac{(p-1)~{_3}F_2(1+p,1+p,1+q;1,p-1;\frac{\bar{n}}{1+\bar{n}})}
{{_3}F_2(1+p,1+p,1+q;1,p;\frac{\bar{n}}{1+\bar{n}})}\\
& & -\frac{p~{_3}F_2(1+p,1+p,1+q;1,p;\frac{\bar{n}}{1+\bar{n}})}{{_2}F_1(1+q,1+p;1;\frac{\bar{n}}{1+\bar{n}})}
\end{eqnarray}

\begin{figure}[h]
\centering
\subfigure{\includegraphics[width=4cm]{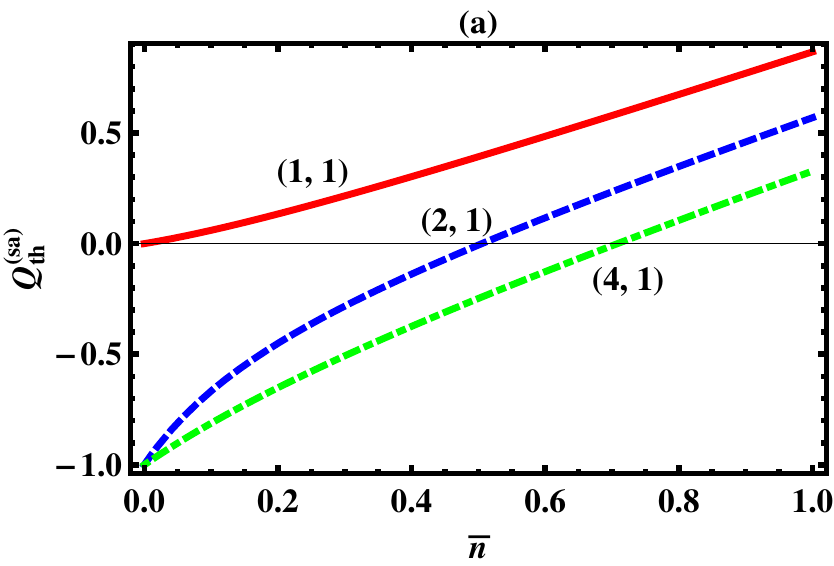}}\hspace{0.5cm}
\subfigure{\includegraphics[width=4cm]{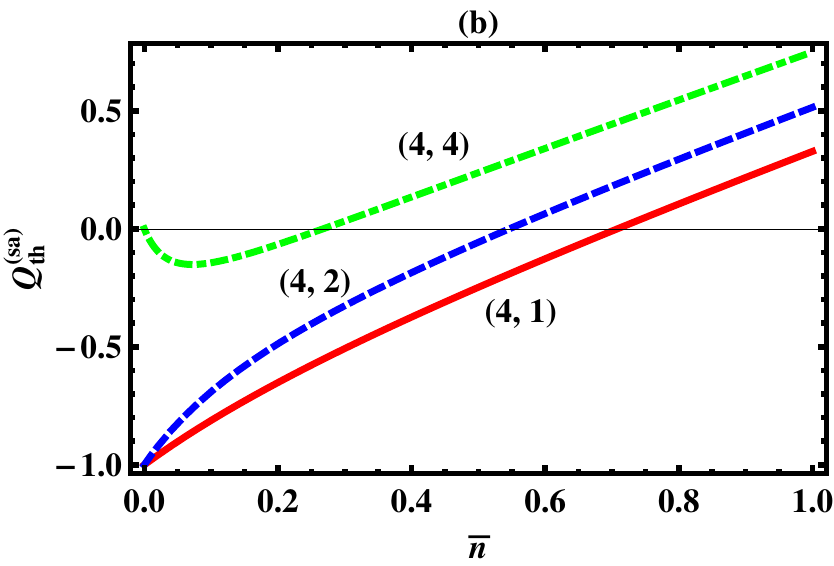}}
\subfigure{\includegraphics[width=4cm]{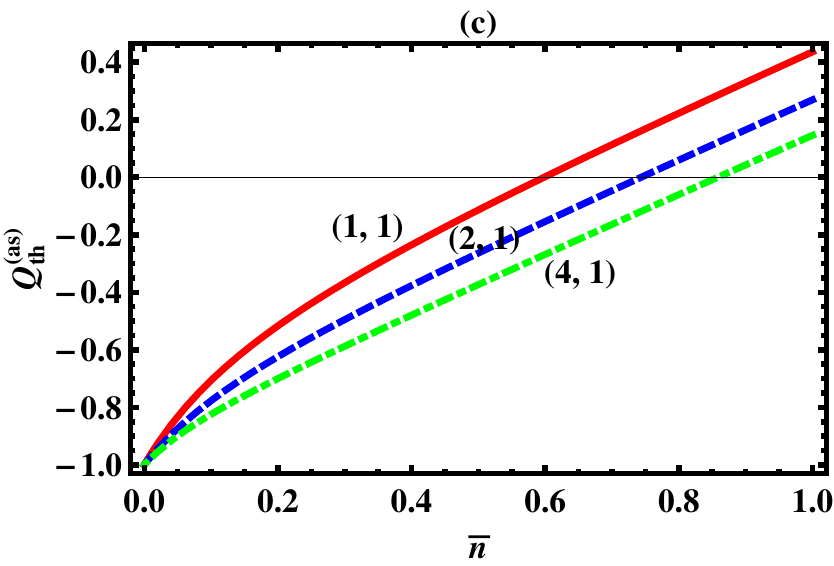}}\hspace{0.5cm}
\subfigure{\includegraphics[width=4cm]{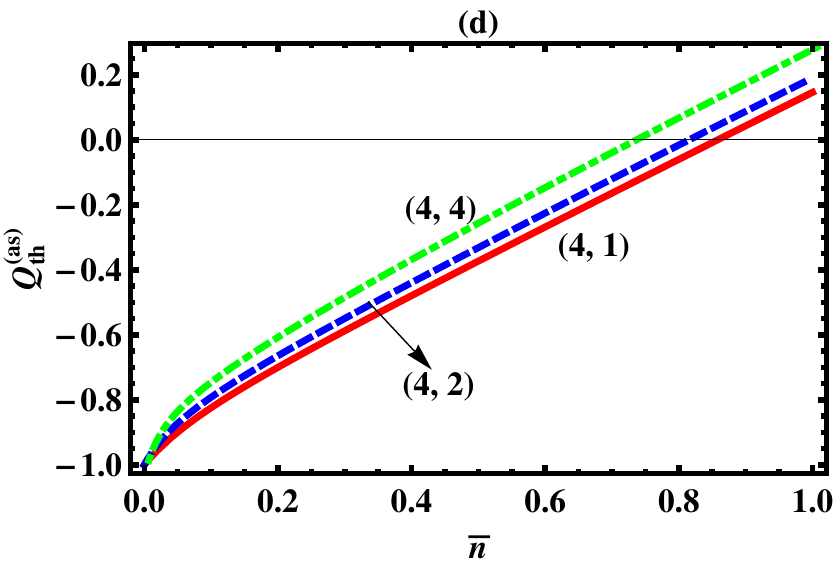}}

\caption{(Color online) Mandel's $Q$ parameter for photon-added-then-subtracted thermal state (upper row) and photon-subtracted-then-added thermal state (lower row) as a function of mean photon number with different $(p,~q)$'s.}
\label{fig3}
\end{figure}
In order to see the variation of the $Q$ parameter against the mean photon number $\bar{n}$, Mandel's $Q$ is plotted as a function of $\bar{n}$ in Fig.~\ref{fig3}. In the range of $(-1,~1/2)$, $Q$ increases monotonically with $\bar{n}$, regardless of several $(p,~q)$ values. Both the added-then-subtracted and subtracted-then-added $Q$ curves are partially negative and partially positive which indicates that the fields enjoying sub-Poissonian characteristic also obey super-Poissonian distribution after a certain limit of $\bar{n}$. For fixed $q$, as depicted in Fig.~\ref{fig3}(a), the values of $Q_{\text{th}}^{(sa)}$ decrease as $p$ increases. But in Fig.~\ref{fig3}(b), where the number of addition is fixed and the number of subtraction is varied, $Q$ increases as $q$ increases. It seems that increasing creation (annihilation) number may produce stronger (weaker) sub-Poissonian statistics. On the contrary, Fig.~\ref{fig3}(c) and Fig.~\ref{fig3}(d) have made it clear that adding more photons from a subtracted-then-added thermal field (keeping $q$ fixed) creates a weaker super-Poissonian distribution and subtracting more photons from a subtracted-then-added thermal field (keeping $p$ fixed) creates a stronger super-Poissonian distribution. It is to be pointed out that in general $Q_{\text{th}}^{(sa)}$ reaches Poissonian level ($Q=0$) more rapidly than $Q_{\text{th}}^{(as)}$.

\section{Even Coherent State}
\label{sec4}

The even coherent state [ECS] is defined as a superposition of two coherent states as \cite{dodonov74}
\begin{eqnarray}
|\psi\rangle_{\text{ECS}} = \frac{1}{(2+2e^{-2|\alpha|^2})^{1/2}}(|\alpha\rangle+|-\alpha\rangle),
\label{eq19}
\end{eqnarray}
where $|-\alpha\rangle$ has the same amplitude as $|\alpha\rangle$ but with a phase shift of $\pi$. When $|\alpha|$ is as small as 2, $|\langle \alpha|-\alpha\rangle|^2\approx0$ \cite{jeong05}. Assuming (\ref{eq19}) as the initial state, the density operators for the photon-added-then-subtracted and photon-subtracted-then-added even coherent states are respectively
\begin{eqnarray}
\hat{\rho}_{\text{ECS}}^{(sa)} = N_3a^q a^{\dag p} \hat{\rho}_{\text{ECS}} a^p a^{\dag q},
\label{eq20}
\end{eqnarray}
and
\begin{eqnarray}
\hat{\rho}_{\text{ECS}}^{(as)} = N_4 a^{\dag p}a^q \hat{\rho}_{\text{ECS}} a^{\dag q}a^p,
\label{eq21}
\end{eqnarray}
where $N_3$ and $N_4$ are the normalization constants. For deriving $N_3$ and $N_4$, we use some normal- (antinormal-) ordered operator identities such as  \cite{hong02}
$$a^{\dag p}a^q=:H_{p, q}(a^\dag, a):,\,a^q a^{\dag p}=(-i)^{p+q}:H_{p, q}(ia^\dag, ia):,$$ and
\begin{eqnarray}\nonumber
& & :H_{p, q}(ia^\dag, ia)::H_{u, v}(ia^\dag, ia):\\\nonumber
& = & \sum_{n=0}^{min(p, v)}\frac{p!v!}{n!(p-n)!(v-n)!}:H_{p+u-n, q+v-n}(ia^\dag, ia):,
\end{eqnarray}
where $::$ stands for normal ordering, $H$ is the two-variable Hermite polynomial defined as
\begin{eqnarray}\nonumber
H_{m, n}(x, y) = \sum_{l=0}^{min(m, n)}(-1)^l\frac{m! n!}{l!(m-l)!(n-l)}x^{m-l}y^{n-l}.
\end{eqnarray}
With the help of the above properties and the well-known relation between the bivariate Hermite polynomial and the Laguerre polynomial \cite{wang11}, i.e. $H_{m, m}(x, y) = (-1)^m m! L_m(xy)$,
$N_3$ and $N_4$ can be calculated as
\begin{equation}
\left.
\begin{array}{lcl}
N_3 = \frac{\left(1+e^{-2|\alpha|^2}\right)}{\sum_{m=0}^q \frac{(q!)^2(p+q-m)!}{(-1)^m m!((q-m)!)^2}
L_{p+q-m}^{\text{(sup)}}(|\alpha|^2)},\\\\
N_4 = (-1)^{p+q}N_3,
\end{array}
\right\}
\end{equation}
where $L_{p+q-m}^{\text{(sup)}}(|\alpha|^2)=L_{p+q-m}(|\alpha|^2)+L_{p+q-m}(-|\alpha|^2)$.

\subsection{Photon Number Distribution}

We can find $n$ number of photons in states (\ref{eq20}) and (\ref{eq21}) respectively with the probabilities
\begin{equation}
p_{\text{ECS}}^{(sa)}(n) =
\left\{
\begin{array}{lcl}
\frac{2N_3 e^{-|\alpha|^2}}{(1+e^{-2|\alpha|^2})}\frac{((n+q)!)^2(|\alpha|^2)^{n-p+q}}{n!((n-p+q)!)^2},~~\text{$n-p+q$ even}\\\\
0~~~~~~~~~~~~~~~~~~~~~~~~~~~~~~~~~~~~,~~\text{$n-p+q$ odd}
\end{array}
\right.
\end{equation}
and
\begin{equation}
p_{\text{ECS}}^{(as)}(n) =
\left\{
\begin{array}{lcl}
\frac{2N_4 e^{-|\alpha|^2}}{(1+e^{-2|\alpha|^2})}\frac{n!(|\alpha|^2)^{n-p+q}}{((n-p)!)^2},~~\text{$n-p+q$ even}\\\\
0~~~~~~~~~~~~~~~~~~~~~~~~~~~~,~~\text{$n-p+q$ odd}
\end{array}
\right.
\end{equation}

\begin{figure*}[ht]
\centering
\includegraphics[width=5cm]{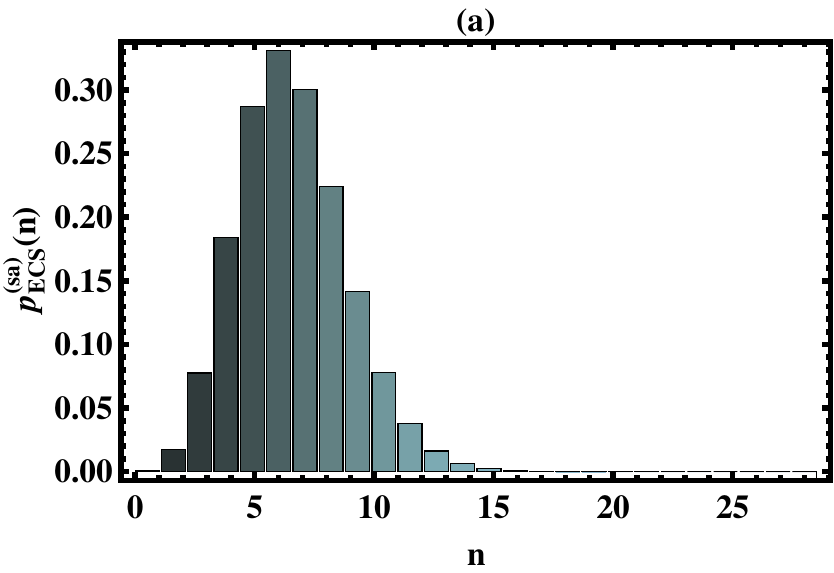}\hspace{1cm}
\includegraphics[width=5cm]{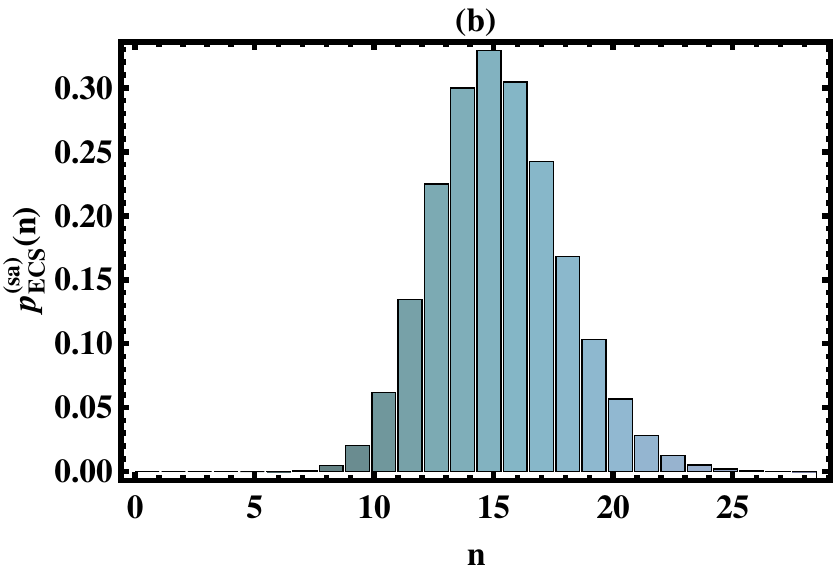}\hspace{1cm}
\includegraphics[width=5cm]{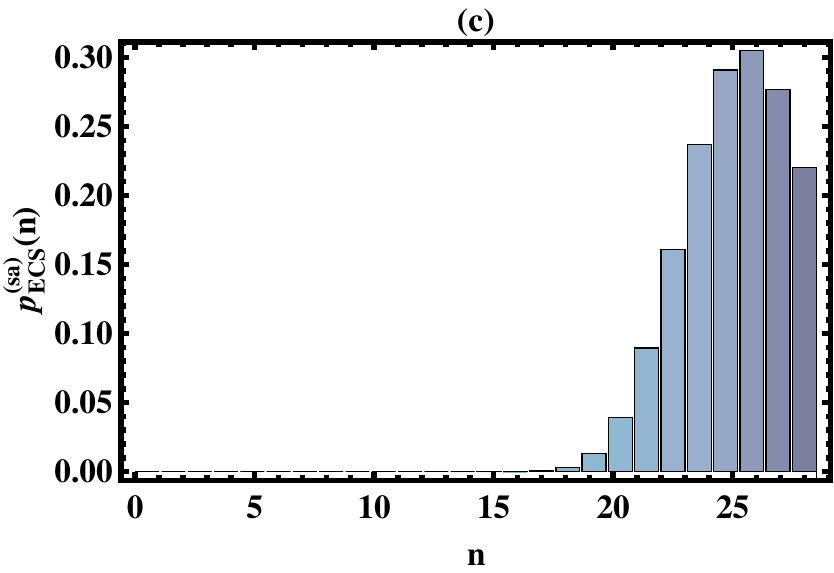}\hspace{1cm}
\includegraphics[width=5cm]{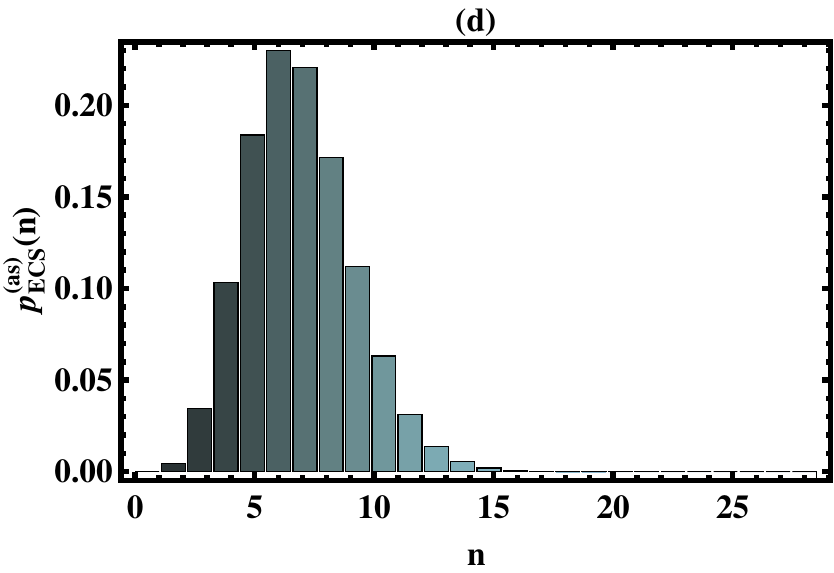}\hspace{1cm}
\includegraphics[width=5cm]{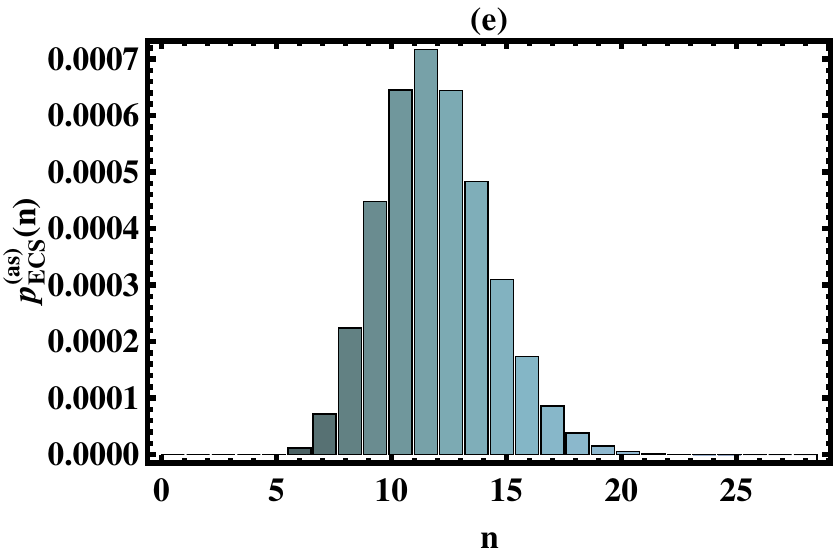}\hspace{1cm}
\includegraphics[width=5cm]{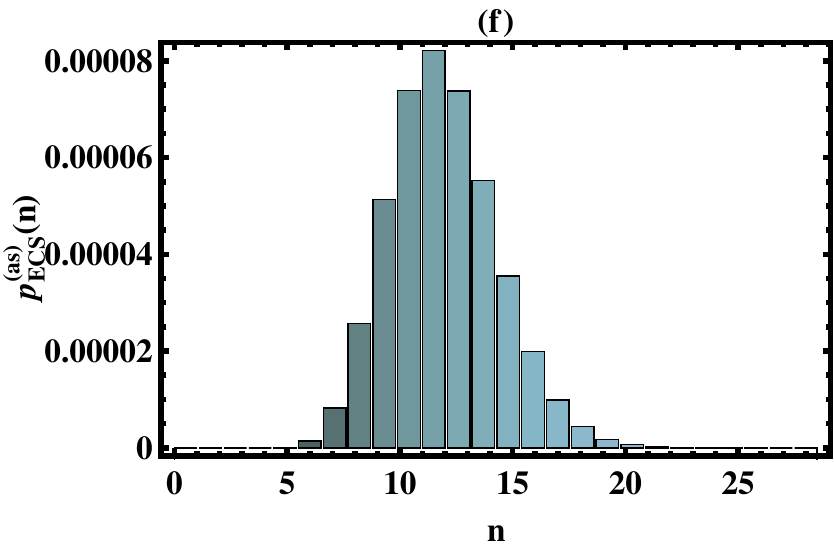}

\caption{(Color online) Photon number distribution of photon-added-then-subtracted (upper line) and photon-subtracted-then-added (lower line) even coherent state is plotted against $n$ for $|\alpha|^2=4$ and (a) $p=q=1$, (b) $p=8$, $q=4$, (c) $p=16$, $q=4$, (d) $p=q=1$, (e) $p=4$, $q=8$ and (f) $p=4$, $q=12$.}
\label{fig4}
\end{figure*}

In Fig.~\ref{fig4}, we examine how the changes in $(p, q)$ affect the photon-added-then-subtracted (photon-subtracted-then-added) even coherent state, when $n-p+q$ is even and $|\alpha|^2=4$. In general, $p_{\text{ECS}}^{(sa)}$ has a broader distribution than $p_{\text{ECS}}^{(as)}$. Figs.~\ref{fig4}(b) and (c) show that by increasing the excitation number one can move the peak towards right for the added-then-subtracted state. But for the subtracted-then-added even coherent state, the increase in $q$ (whenever $q$ differs a lot from $p$) has no significant effect on the position of the peak [see Figs.~\ref{fig4}(e)-(f)].

\subsection{Wigner Distribution}

To find out the analytical expressions for the Wigner functions of added-then-subtracted and subtracted-then-added even coherent states, we recall some basic properties of two-variable Hermite polynomial \cite{wang11}
\begin{eqnarray}\nonumber
& & H_{m, n}(x, y) = \frac{\partial^{m+n}}{{\partial u}^m{\partial v}^n}\exp(-uv+ux+vy)|_{u,v=0},\\\nonumber
& & \frac{\partial^r}{{\partial x}^r}H_{m, n}(x, y) = \frac{m!}{(m-r)!}H_{m-r, n}(x, y),
\end{eqnarray}
and an integral formula \cite{puri01}
\begin{eqnarray}\nonumber
& & \int{\frac{d^2z}{{\pi}^2}\exp(a|z|^2+bz+cz^*+d{z}^2+ez^{*^2})}\\\nonumber
& = & \frac{1}{\sqrt{a^2-4de}}\exp{\left(\frac{-abc+b^2e+c^2d}{a^2-4de}\right)},
\end{eqnarray}
whose convergent condition is $\text{Re}(a\pm d\pm e)<0$ and $\text{Re}(\frac{a^2-4de}{a\pm d\pm e})<0$. Insertion of these formulas into (\ref{eq12})
gives us
\begin{widetext}
\begin{eqnarray}\nonumber
W_{\text{ECS}}^{(sa)}(\beta, \beta^*) & = & \frac{N_3}{(1+e^{-2|\alpha|^2})}\sum_{n=0}^{p} \frac{(-1)^n (p!)^2}{n!((p-n)!)^2}
\left\{\left(|H_{p-n, q}[i(2\beta-\alpha), i\alpha^*]|^2 e^{-2|\alpha-\beta|^2}+|H_{p-n, q}[i(2\beta+\alpha), -i\alpha^*]|^2 e^{-2|\alpha+\beta|^2}\right)\right.\\\nonumber
& & \left.+e^{-2|\beta|^2}\left(H_{p-n, q}[i(2\beta-\alpha), -i\alpha^*]\overline{H_{p-n, q}[i(2\beta+\alpha), i\alpha^*]}e^{2(\alpha\beta^*-\alpha^*\beta)}\right.\right.\\
& & \left.\left.+\overline{H_{p-n, q}[i(2\beta-\alpha), -i\alpha^*]}H_{p-n, q}[i(2\beta+\alpha), i\alpha^*]e^{2(\alpha^*\beta-\alpha\beta^*)}\right)\right\},
\label{eq25}
\end{eqnarray}
\end{widetext}
and
\begin{widetext}
\begin{eqnarray}\nonumber
W_{\text{ECS}}^{(as)}(\beta, \beta^*) & = & \frac{N_4}{(1+e^{-2|\alpha|^2})}\sum_{n=0}^{p} \frac{(-1)^n (p!)^2}{n!((p-n)!)^2}
\left\{\left(|H_{p-n, q}[2\beta-\alpha, \alpha^*]|^2 e^{-2|\alpha-\beta|^2}+|H_{p-n, q}[2\beta+\alpha, -\alpha^*]|^2 e^{-2|\alpha+\beta|^2}\right)\right.\\\nonumber
& & \left.+e^{-2|\beta|^2}\left(H_{p-n, q}[2\beta-\alpha, -\alpha^*]\overline{H_{p-n, q}[2\beta+\alpha, \alpha^*]}e^{2(\alpha\beta^*-\alpha^*\beta)}\right.\right.\\
& & \left.\left.+\overline{H_{p-n, q}[2\beta-\alpha, -\alpha^*]}H_{p-n, q}[2\beta+\alpha, \alpha^*]e^{2(\alpha^*\beta-\alpha\beta^*)}\right)\right\},
\label{eq26}
\end{eqnarray}
\end{widetext}
where $\overline{H_{m, n}(x, y)}$ denotes complex conjugate of $H_{m, n}(x, y)$. The partial negativity of Wigner function is a clear signature of nonclassical character of the related state. But this condition is one-sided i.e. one cannot conclude the state is classical when the Wigner function is positive everywhere. For example, the Wigner function of the squeezed state is Gaussian and positive everywhere but it is a well-known nonclassical state.
\begin{figure*}[ht]
\centering
\includegraphics[width=5cm]{fig11_wig_ECS.pdf}\hspace{1cm}
\includegraphics[width=5cm]{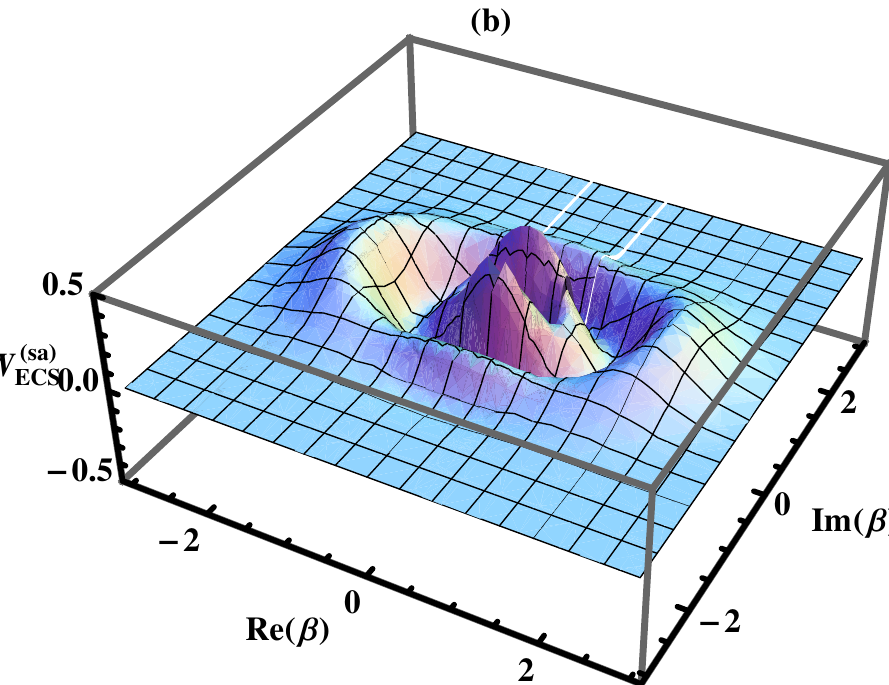}\hspace{1cm}
\includegraphics[width=5cm]{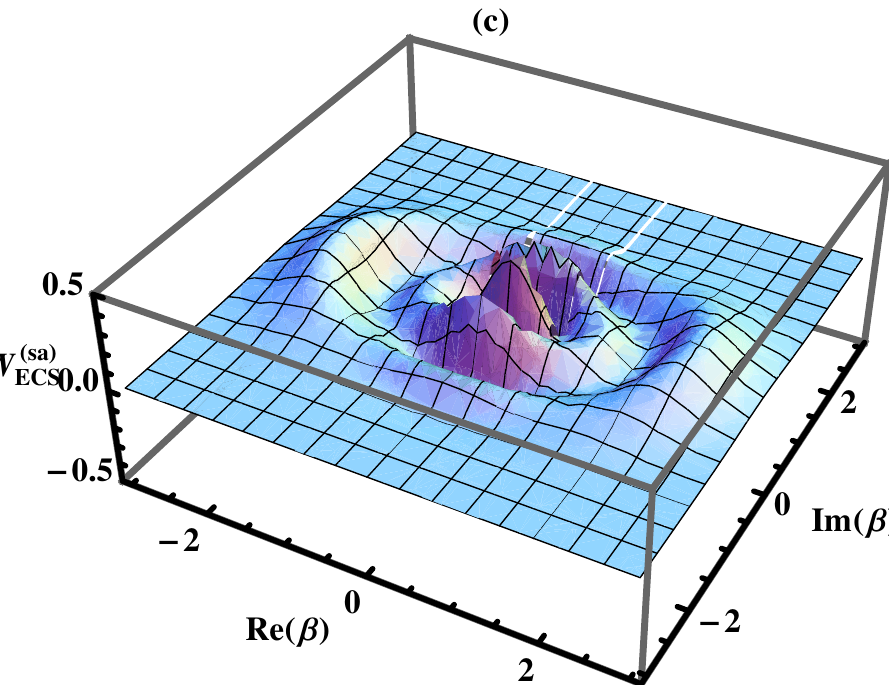}\hspace{1cm}
\includegraphics[width=5cm]{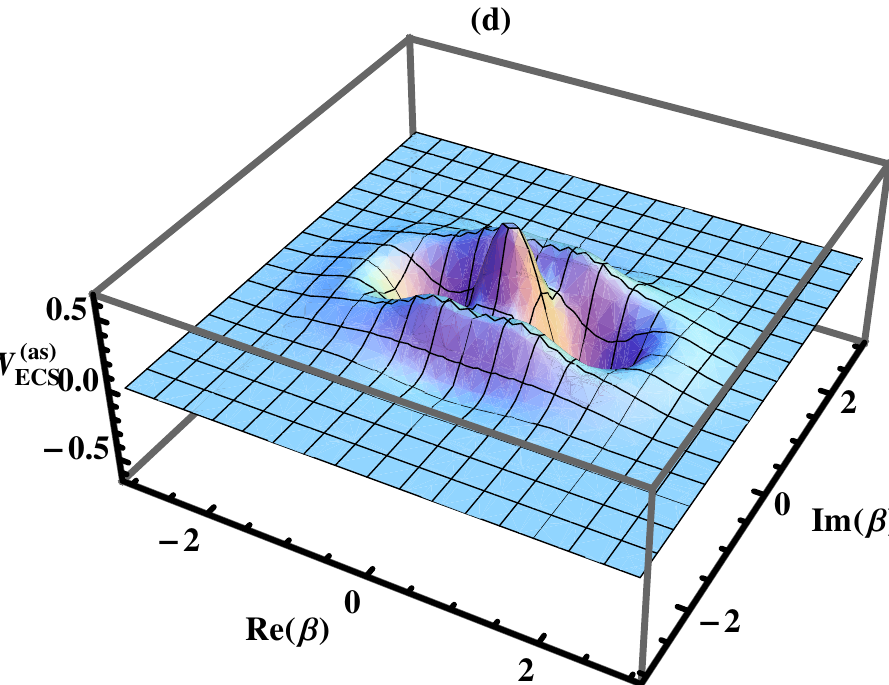}\hspace{1cm}
\includegraphics[width=5cm]{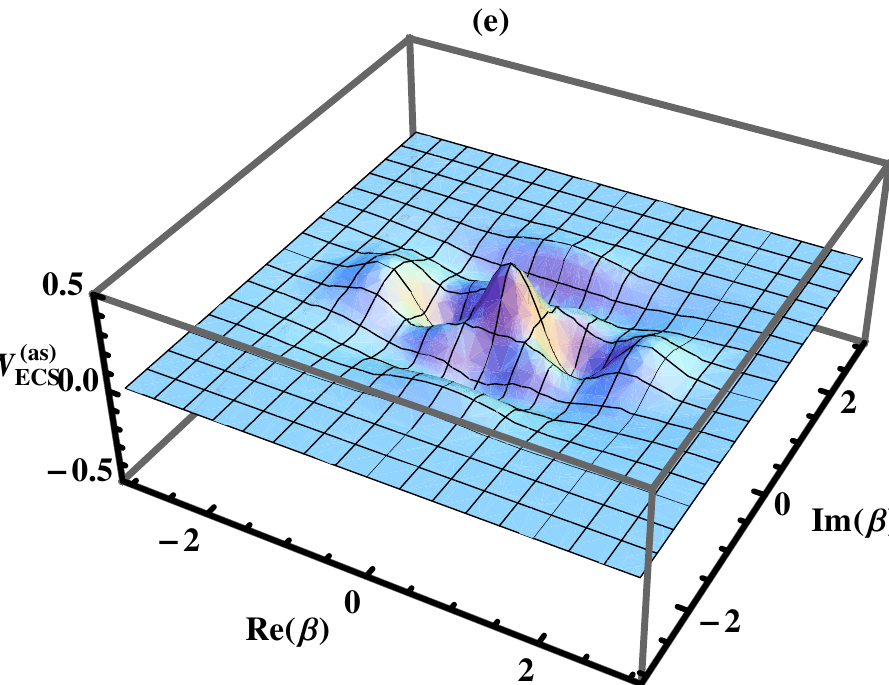}\hspace{1cm}
\includegraphics[width=5cm]{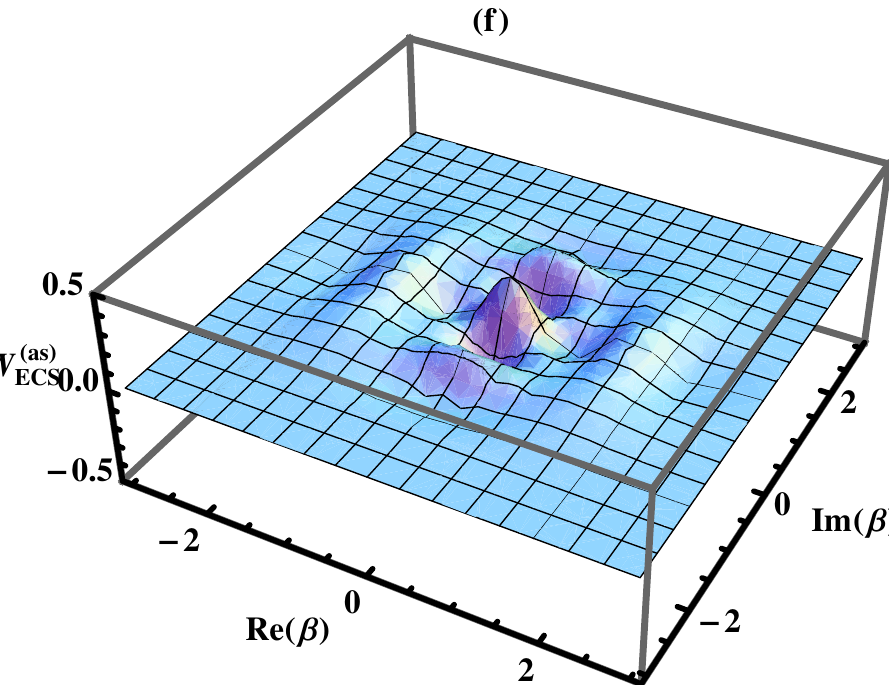}

\caption{(Color online) Wigner function of photon-added-then-subtracted (upper row) and photon-subtracted-then-added (lower row) even coherent state as a function of Re$(\beta)$ and Im$(\beta)$ for $\alpha=1$ and different $(p,q)$'s (a) $(1,1)$, (b) $(2,1)$, (c) $(3,1)$, (d) $(1,1)$, (e) $(2,1)$ and (f) $(3,1)$.}
\label{fig5}
\end{figure*}

Based on Eqs.~(\ref{eq25}) and (\ref{eq26}), we plot the Wigner function in phase space with several combinations of $(p,q)$ and $\alpha=1$. The nonclassical character of both the added-then-subtracted and subtracted-then-added even coherent states are depicted in Fig.~\ref{fig5}. In case of $p=q=1$, the distribution $W_{\text{ECS}}^{(sa)}(\beta, \beta^*)$ almost coincides with the distribution $W_{\text{ECS}}^{(as)}(\beta, \beta^*)$. We observe that the central Gaussian peak of the added-then-subtracted even coherent state first transforms to a single peak with a deep crater and then again a single Gaussian peak comes out of this crater as $p$ changes from 1 to 2 [Fig.~\ref{fig5}(b)] and 2 to 3 [Fig.~\ref{fig5}(c)]. While for subtracted-then-added even coherent state, the bumps at the two ends of the $x$-axis slowly disappear with $p$. We further notice that the partial negative region of the added-then-subtracted (subtracted-then-added) even coherent state gradually diminishes with increasing $p$. This implies that increasing photon addition number causes the lose of nonclassicality of the state. In fact if we increase $q$ together with $p$, $W_{\text{ECS}}^{(as)}(\beta, \beta^*)$ just reduces to show a nearly plane region.

\begin{figure}
\centering
\subfigure[~$p=1$]
{\includegraphics[width=6cm]{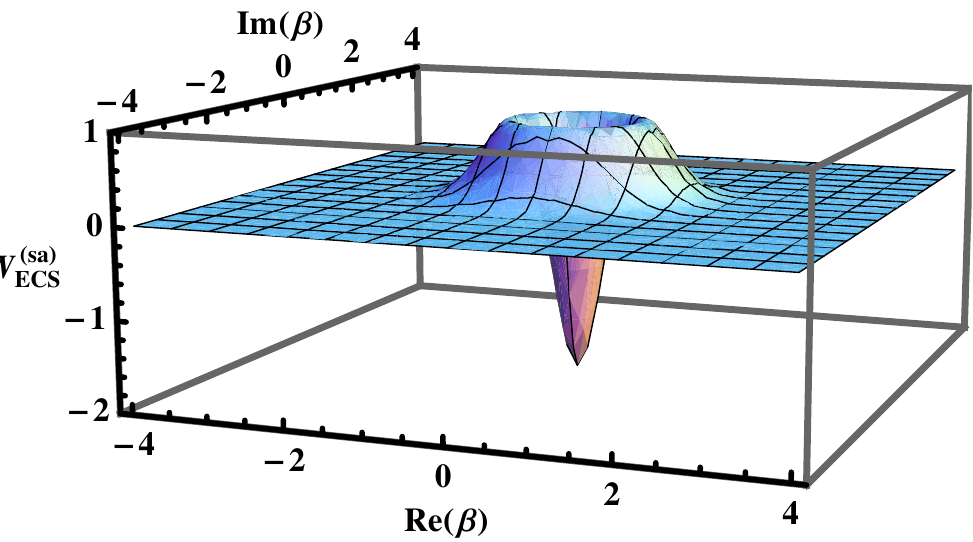}}
\subfigure[~$p=5$]
{\includegraphics[width=6cm]{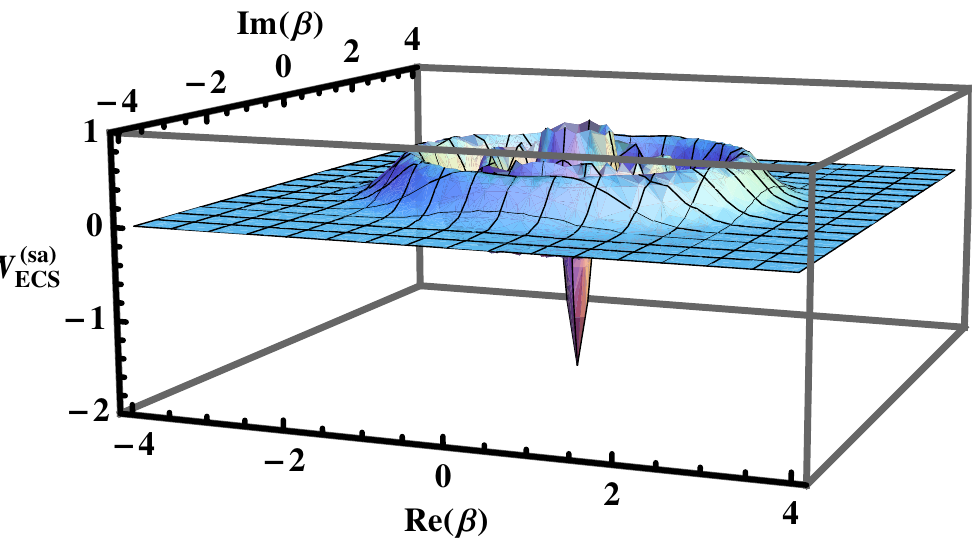}}

\caption{(Color online) Wigner function of only photon-added even coherent state with $\alpha=0.1$.}
\label{fig6}
\end{figure}
In Fig.~\ref{fig6}, we show Wigner functions of only photon-added even coherent state for $\alpha=0.1$. Choosing (a) $p=1$ and (b) $p=5$, we obtain partially negative Wigner functions which look like those presented in \cite{dodonov96}.

\subsection{Mandel's $Q$ Parameter}

Mandel's $Q$ parameter for even coherent state can be derived using the following relations:

\begin{eqnarray}\nonumber
{\langle a^\dag a\rangle}^{(sa)} & = & \frac{N_3}{(1+e^{-2|\alpha|^2})}\sum_{m=0}^{q+1}\frac{((q+1)!)^2(p+q+1-m)!}{(-1)^m m!((q+1-m)!)^2}\\
& & \times L_{p+q+1-m}^{\text{(sup)}}(|\alpha|^2),
\label{eq27}
\end{eqnarray}
\begin{eqnarray}\nonumber
{\langle a^{\dag 2} a^2\rangle}^{(sa)} & = & \frac{N_3}{(1+e^{-2|\alpha|^2})}\sum_{m=0}^{q+2}\frac{((q+2)!)^2(p+q+2-m)!}{(-1)^m m!((q+2-m)!)^2}\\
& & \times L_{p+q+2-m}^{\text{(sup)}}(|\alpha|^2),
\label{eq28}
\end{eqnarray}
\begin{widetext}
\begin{eqnarray}\nonumber
{\langle a^\dag a\rangle}^{(as)} = \frac{N_4}{(1+e^{-2|\alpha|^2})}\sum_{m=0}^{q}\frac{(-1)^{p+q+1-m}(q!)^2(p+q-m)!}{m!((q-m)!)^2}
\left[(p+q+1-m)L_{p+q+1-m}^{\text{(sup)}}(|\alpha|^2)+L_{p+q-m}^{\text{(sup)}}(|\alpha|^2)\right],\\
\label{eq29}
\end{eqnarray}
\end{widetext}
and
\begin{widetext}
\begin{eqnarray}\nonumber
{\langle a^{\dag 2} a^2\rangle}^{(as)} & = & \frac{N_4}{(1+e^{-2|\alpha|^2})}\sum_{m=0}^{q}\frac{(-1)^{p+q+2-m}(q!)^2(p+q-m)!}{m!((q-m)!)^2}
\left[(p+q+2-m)(p+q+1-m)L_{p+q+2-m}^{\text{(sup)}}(|\alpha|^2)\right.\\
& & \left.+4(p+q+1-m)L_{p+q+1-m}^{\text{(sup)}}(|\alpha|^2)+2L_{p+q-m}^{\text{(sup)}}(|\alpha|^2)\right].
\label{eq30}
\end{eqnarray}
\end{widetext}
Substituting (\ref{eq27}) and (\ref{eq28}) into (\ref{eq16}) and (\ref{eq29}) and (\ref{eq30}) again into (\ref{eq16}) we determine $Q_{\text{ECS}}^{(sa)}$ and $Q_{\text{ECS}}^{(as)}$ respectively.

\begin{figure}[ht]
\centering
\subfigure{\includegraphics[width=4cm]{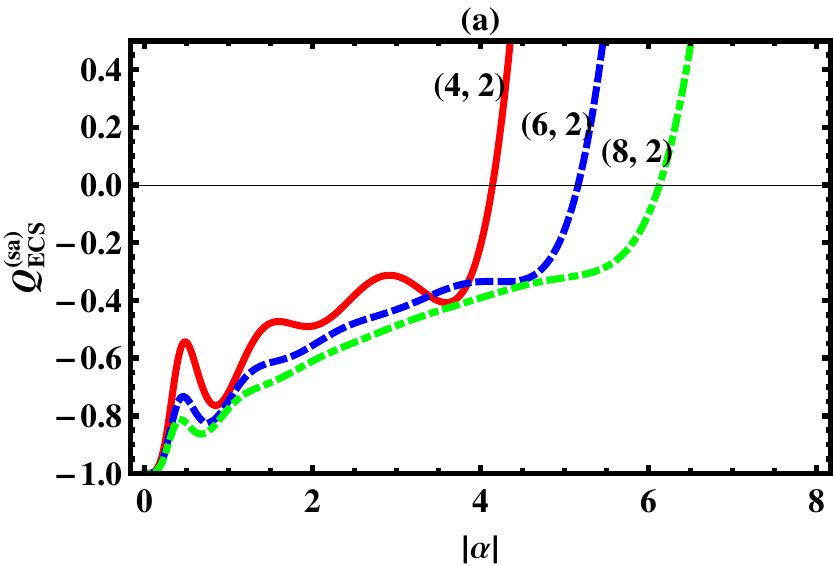}}\hspace{0.5cm}
\subfigure{\includegraphics[width=4cm]{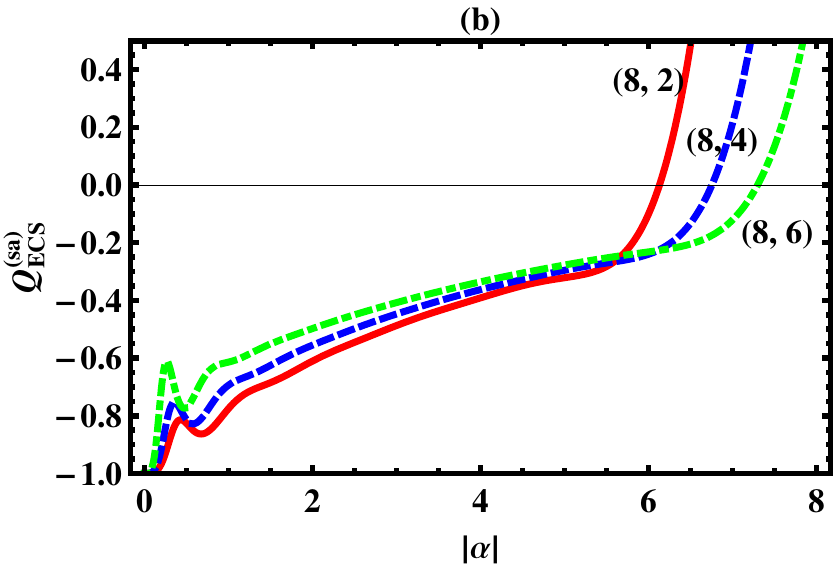}}
\subfigure{\includegraphics[width=4cm]{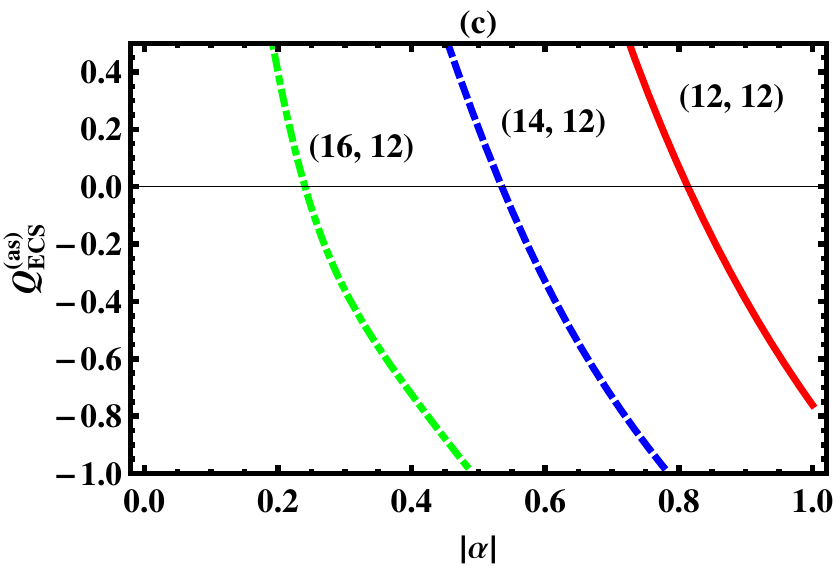}}\hspace{0.5cm}
\subfigure{\includegraphics[width=4cm]{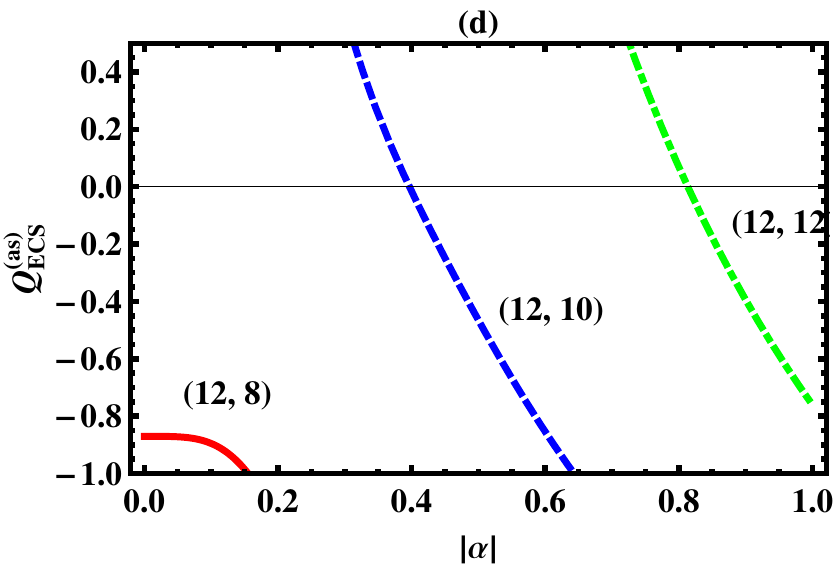}}

\caption{(Color online) Mandel's $Q$ parameter for photon-added-then-subtracted even coherent state (upper row) and photon-subtracted-then-added even coherent state (lower row) as a function of $|\alpha|$  with different $(p,~q)$'s.}
\label{fig7}
\end{figure}
Fig.~\ref{fig7} clearly shows the changes of $Q_{\text{ECS}}^{(sa)}$ and $Q_{\text{ECS}}^{(as)}$ curves for a fixed number of one operation and a variable number of other operation. Fig.~\ref{fig7}(a) and Fig.~\ref{fig7}(b) respectively exhibit that $Q_{\text{ECS}}^{(sa)}$ goes far away from Poissonian level as $p$ increases, keeping $q$ fixed and comes closer to Poissonian level as $q$ increases, keeping $p$ fixed. The values of $Q_{\text{ECS}}^{(as)}$ also conclude the same. For increasing photon creation number, $Q_{\text{ECS}}^{(sa)}$ sticks to its sub-Poissonian character much more but for increasing photon subtracting number, $Q_{\text{ECS}}^{(as)}$ changes its characteristic to indicate super-Poissonian distribution.

\section{Conclusion}
\label{sec5}

In this article, we have introduced the density matrices of different quantum states after applying an operator combination $a^q a^{\dag p}$ or $a^{\dag p}a^q$ to them. Assuming the field to be initially either in a thermal state or in an even coherent state, we have investigated the statistical properties depending on the analytical expression of the normalization constant, photon number distribution, Wigner function and Mandel's $Q$ parameter. Two different criteria i.e. negativity of Wigner function and Poissonian statistics of Mandel's $Q$ have been used to reveal the nonclassicality of photon-added-then-subtracted (thermal or even coherent) state and photon-subtracted-then-added (thermal or even coherent) state. We have noticed that the Wigner function of thermal state has no negative region at all but the Wigner function of even coherent state exhibits a partial negative region in phase space which is a clear evidence for the nonclassicality of the state.
In case of even coherent state, as $(p,q)$ becomes larger, $W_{\text{ECS}}^{(sa)}(\beta, \beta^*)$ and $W_{\text{ECS}}^{(as)}(\beta, \beta^*)$ change in an almost opposite way. In addition, the $Q$ parameter of thermal state presenting negative values becomes $>0$ after a certain limit of $\bar{n}$. We have seen also irrespective of $(p, q)$ values, $Q_{\text{ECS}}^{(sa)}$ represents super-Poissonian curves after a certain value of $|\alpha|$. In conclusion, the different results for the different orders of adding and subtracting multiphotons to the initial state clearly proves the non-commutativity between $a$ and $a^\dag$.

\begin{center}
\textbf{ACKNOWLEDGEMENT}
\end{center}
AC thanks National Board of Higher Mathematics, Department of Atomic Energy, India for the financial support.


\begin{thebibliography}{99}

\bibitem{agarwal91} G. S. Agarwal and K. Tara, Phys. Rev. A \textbf{43}, 492 (1991).
\bibitem{ban96} M. Ban, J. Mod. Opt. \textbf{43}, 1281 (1996).
\bibitem{dakna98} M. Dakna, L. Kn\"{o}ll and D.-G. Welsch, Euro. Phys. J. D \textbf{3}, 295 (1998).
\bibitem{sun08} Q. Sun, M. Al-Amri and M. S. Zubairy, Phys. Rev. A \textbf{78}, 043801 (2008).
\bibitem{parigi07} V. Parigi, A. Zavatta, M. S. Kim and M. Bellini, Science \textbf{317}, 1890 (2007).
\bibitem{kim08} M. S. Kim, H. Jeong, A. Zavatta, V. Parigi and M. Bellini, Phys. Rev. Lett. \textbf{101}, 260401 (2008).
\bibitem{jones97} G. N. Jones, J. Haight and C. T. Lee, Quantum Semiclass. Opt. \textbf{9}, 411 (1997).
\bibitem{barbieri10} M. Barbieri \textit{et al}., Phys. Rev. A \textbf{82}, 063833 (2010).
\bibitem{ourjoumtsev07} A. Ourjoumtsev, A. Dantan, R. Tualle-Brouri and P. Grangier, Phys. Rev. Lett. \textbf{98}, 030502 (2007).
\bibitem{xiang10} G. Y. Xiang, T. C. Ralph, A. P. Lund, N. Walk and G. J. Pryde, Nature Photon \textbf{4}, 316 (2010).
\bibitem{zavatta04} A. Zavatta, S. Viciani and M. Bellini, Science \textbf{306}, 660 (2004).
\bibitem{zavatta05} A. Zavatta, S. Viciani and M. Bellini, Phys. Rev. A \textbf{72}, 023820 (2005).
\bibitem{zavatta07} A. Zavatta, V. Parigi and M. Bellini, Phys. Rev. A \textbf{75}, 052106 (2007).
\bibitem{scully97} M. O. Scully and M. S. Zubairy, Quantum Optics, Cambridge University Press, (1997).
\bibitem{braunstein05} S. L. Braunstein and P. van Loock, Rev. Mod. Phys. \textbf{77}, 513 (2005).
\bibitem{marek08} P. Marek, H. Jeong and M. S. Kim, Phys. Rev. A \textbf{78}, 063811 (2008).
\bibitem{kim41} M. S. Kim, J. Phys. B \textbf{41}, 133001 (2008).
\bibitem{yang09} Y. Yang and F.-L. Li, J. Opt. Soc. Am. B. \textbf{26}, 830 (2009).
\bibitem{lee91} C. T. Lee, Phys. Rev. A \textbf{44}, R2775 (1991).
\bibitem{pathak05} P. K. Pathak and G. S. Agarwal, Phys. Rev. A \textbf{71}, 043823 (2005).
\bibitem{allevi10} A. Allevi, A. Andreoni, M. Bondani, M. G. Genoni and S. Olivares, Phys. Rev. A \textbf{82}, 013816 (2010).
\bibitem{mandel79} L. Mandel, Opt. Lett. \textbf{4}, 205 (1979).
\bibitem{dodonov74} V. V. Dodonov, I. A. Malkin and V. I. Man'ko, Physica  \textbf{72}, 597 (1974).
\bibitem{jeong05} H. Jeong and T. C. Ralph, arXiv e-print: quant-ph/0509137v1 (2005).
\bibitem{hong02} H.-Y. Fan and Y. Fan, Commun. Theo. Phys. \textbf{38}, 297 (2002).
\bibitem{wang11} Z. Wang, H.-C. Yuan and H.-Y. Fan, J. Opt. Soc. Am. B \textbf{28}, 1964 (2011).
\bibitem{puri01} R. R. Puri, Mathematical Methods of Quantum Optics, Springer-Verlag, (2001).
\bibitem{dodonov96} V. V. Dodonov, Ya. A. Korennoy, V. I. Man'ko and Y. A. Moukhin, Quantum Semiclass. Opt. \textbf{8}, 413 (1996).

\end{thebibliography}
\end{document}